# Magnetic systems at criticality: different signatures of scaling

R. Pełka[1], P. Konieczny[1], M. Fitta[1], M. Czapla[1], P.M. Zieliński[1], M. Bałanda[1], T. Wasiutyński[1], Y. Miyazaki[2], A. Inaba[2], D. Pinkowicz[3], B. Sieklucka[3]

[1] The H. Niewodniczański Institute of Nuclear Physics Polish Academy of Sciences, Radzikowskiego 152, 31-342 Kraków, Poland

[2] Research Center for Structural Thermodynamics, Graduate School of Science, Osaka University, Toyonaka, Osaka 560-0043, Japan

[3] Faculty of Chemistry, Jagiellonian University, Ingardena 3, 30-060 Kraków, Poland

## Abstract

Different aspects of critical behaviour of magnetic materials are presented and discussed. The scaling ideas are shown to arise in the context of purely magnetic properties as well as in that of thermal properties as demonstrated by magnetocaloric effect or combined scaling of excess entropy and order parameter. Two non-standard approaches to scaling phenomena are described. The presented concepts are exemplified by experimental data gathered on four representatives of molecular magnets.

1. Introduction

Many times we observe that completely different systems exhibit the same physics. Such physics is said to be universal and its most famous example is the critical phenomena [1,2]. In the vicinity of the second-order phase transitions the correlation length diverges and the microscopic details become unimportant. The corresponding array of phenomena observed there are characterized only by a few ingredients: dimensionality, interaction range and symmetry of the order parameter. The corresponding state of matter is referred to as being critical. Accordingly, fluids and magnets exhibit the same critical-point exponents describing the quantitative nature of the observed singularities. The universality in critical phenomena has been one of the central issues in condensed matter physics.

In the present paper we focus on the critical behaviour of a selected class of condensed matter systems, i.e. magnetic materials. Within this broad class of materials we exclusively concentrate on molecular magnets whose structural elements exhibit localized magnetic moments. There are two main reasons for our choice. On the one hand, the localization of constituent magnetic moments is a feature making a direct reference to the existing models based on discrete spin-like degrees of freedom like the XY model, the Heisenberg model or the Ising model. On the other hand, molecular magnets represent materials which have been studied by our group for more than 10 years now providing a unique opportunity for a detailed insight into their critical behaviour. Apart from the standard analysis of critical phenomena we had the chance to come across some more exotic types of scaling behaviour like, e.g., combined scaling of excess entropy and order parameter. The present contribution is thought to collect and summarize our experience with the critical behaviour in molecular magnets, which may come in useful for anybody willing to study magnetic systems at criticality.

To illustrate the different aspects of the scaling behaviour four representatives of molecular magnets were selected. Compound **1** with the formula $\{[Mn^{II}(pydz)(H_2O)_2][Mn^{II}(H_2O)_2][Nb^{IV}(CN)_8]\cdot 2H_2O\}_n$, where "pydz" stands for pyridazine ($C_4H_4N_2$), crystallizes in the monoclinic space group $P2_1/c$ [3,4]. The compound is a three-dimensional (3D) cyanido-bridged framework composed of corrugated square-grid motifs parallel to the *bc* crystallographic plane and involving alternating arrangement of $Mn^{II}$ and $Nb^{IV}$ centres. These are linked at the $Nb^{IV}$ centres by the ladder motifs running along the *a* crystallographic axis. It exhibits the transition to a magnetically ordered state at $T_c \approx 42$ K. The most striking physical property of this compound is a two-step shift of the ordering temperature toward higher temperatures observed upon dehydration. The first stage of dehydration involving the removal of the crystallization water and two aqua ligands enhances the intranetwork exchange interaction and triggers the increase of magnetic ordering temperature up to 68 K. Further dehydration leading to the removal of the remaining two water molecules affects the increase of the critical temperature up to about 100 K. This anhydrous compound with the formula $\{[Mn^{II}_2(pydz)][Nb^{IV}(CN)_8]\}_n$ is the second representative and will be referred to as **2**. The third representative **3** is a coordination polymer $\{(tetrenH_5)_{0.8}Cu^{II}_4[W^V(CN)_8]_4\cdot 7H_2O\}_n$, (tetren=tetraethylenepentamin) revealing the signatures of the unique Berezinskii-Kosterlitz-Thouless transition at $T_c \approx 33$ K [5-10]. It crystallizes in orthorhombic crystallographic system

(space group *Cmc*2$_1$) and is built of cyanobridged copper-tungsten anionic double-layer sheets lying in the *ac* plane. The space between the double layers is filled with water molecules and tetrenH$_5^{5+}$ solvent molecules. The spin carriers in the system are Cu$^{II}$ (*S*=1/2) and W$^{V}$ (*S*=1/2) ions. The last compound denoted by **4** is {[Fe$^{II}$(pyrazole)$_4$]$_2$[Nb$^{IV}$(CN)$_8$]·4H$_2$O}$_n$ (pyrazole=C$_3$H$_4$N$_2$) [11]. It crystallizes in the tetragonal space group *I*4$_1$/*a* and its structure consists of a 3D cyanido-bridged Fe$^{II}$-NC-Nb$^{IV}$ skeleton decorated with pyrazole molecules coordinated to 3d metal centres. The presence of four and two bridging cyanido ligands per one Nb$^{IV}$ and one Fe$^{II}$ centre, respectively, gives rise to the 4:2 connectivity type, quite unique for 3D coordination systems. This compound reveals the transition to a magnetically ordered phase at $T_c \approx 8$ K.

The paper has been organized as follows. We start with the definitions of critical-point exponents pertinent to a magnetic system in Section 2. In the next Section 3 we go on to discussing an important theoretical aspect of the scaling behaviour, namely the static scaling hypothesis. Section 4 demonstrates a useful tool for finding critical-point exponents proposed by Kouvel and Fisher. Section 5 deals with the generalized Curie-Weiss law and its consequences for critical behaviour in magnetic materials. In Section 6 we discuss the magnetocaloric effect and its scaling properties. Combined scaling of excess entropy and order parameter is presented in Section 7. We wind up in Section 8 by an array of general remarks and conclusions.

2. Definitions of the critical-point exponents

Let us begin with a precise and general definition of a critical-point exponent [12,13] used to describe the behaviour near the critical point of a general function $f(\varepsilon)$, where

$$\varepsilon \equiv \frac{T - T_c}{T_c} \tag{1}$$

serves as a dimensionless variable to measure the distance in temperature from the critical temperature. Assuming that the function $f(\varepsilon)$ is positive and continuous for sufficiently small, positive values of $\varepsilon$, we define the critical point exponent $\varphi$ associated with this function as the following limit

$$\varphi = \lim_{\varepsilon \to 0} \frac{\ln f(\varepsilon)}{\ln \varepsilon} \tag{2}$$

As a shorthand notation we frequently denote the fact that $\varphi$ is the critical point exponent for the function $f(\varepsilon)$ by writing

$$f(\varepsilon) \propto \varepsilon^{\varphi}. \tag{3}$$

Let us stress that the above relation does not imply the relation

$$f(\varepsilon) = F\varepsilon^{\varphi}. \tag{4}$$

In fact, it is relatively rare that the behaviour of a typical thermodynamic function is as simple as Eq. (4). In general, additional correction terms are required, and Eq. (4) is replaced by a functional expression such as

$$f(\varepsilon) = F\varepsilon^{\varphi}(1 + R\varepsilon^{\psi} + \ldots) \quad (\psi > 0). \tag{5}$$

The immediate consequence of the functional behaviour given by Eq. (3) is that a log-log plot of experimental points should display a straight-line behaviour sufficiently near the critical point, and the critical-point exponent is easily determined as the slope of this straight-line region.

Let us start the definitions of particular critical-point exponents related to magnetic media with those associated with magnetization $M(T,H)$ of a system. Critical-point exponent $\beta$ determines the asymptotic behaviour of the zero-field magnetization $M(\varepsilon, H=0)$ near the critical point, thus we may write

$$M(\varepsilon, H=0) \propto (-\varepsilon)^{\beta} \quad \text{where} \quad \varepsilon < 0 \ (T < T_c). \tag{6}$$

If we set $\varepsilon = 0$ ($T = T_c$) and take the limit $H \to 0$, another exponent $\delta$ emerges, defined by the relation

$$M(\varepsilon = 0, H) \propto H^{1/\delta} \tag{7}$$

From a practical point of view there is no magnetometric method that is able to assure an unambiguous determination of the critical-point exponent $\beta$ as they usually require the application of an external magnetic field masking the critical fluctuations. By contrast, the zero-field (ZF) mode of the muon spin rotation (μSR) experiment is perfectly suited to provide precise insight into the temperature dependence of the order parameter through the quasistatic local field values [14-18]. In such an experiment fully spin-polarized positive muons enter a sample, thermalize and stop at sites with a local surplus of the negative electric charge. If there is a nonvanishing local magnetic field at the site not parallel to the magnetic moment of a muon, it starts to precess with a frequency proportional to the magnitude of this local field. The trajectory of the precession is traced by the positron emission registered by the backward and forward detectors. The time evolution of the spin polarization of the implanted muons is detected by measuring the asymmetry function $A(t) = (N_B - N_F)/(N_B + N_F)$, where $N_F$ and $N_B$ denote the numbers of decay positrons emitted forward and backward, respectively. Figure 1 shows the temperature dependence of the local magnetic fields inferred from the ZF µSR experiment on compound **1**.

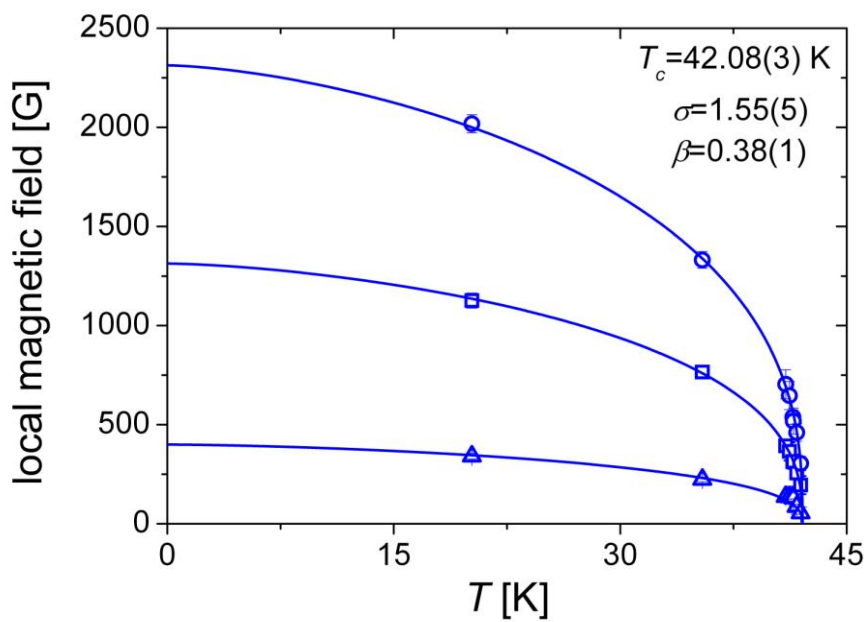

**Fig. 1.** Temperature dependence of the local magnetic field as inferred from the µSR experiment on compound **1**. Three components of the internal field indicate three possible muon stopping sites in the sample.

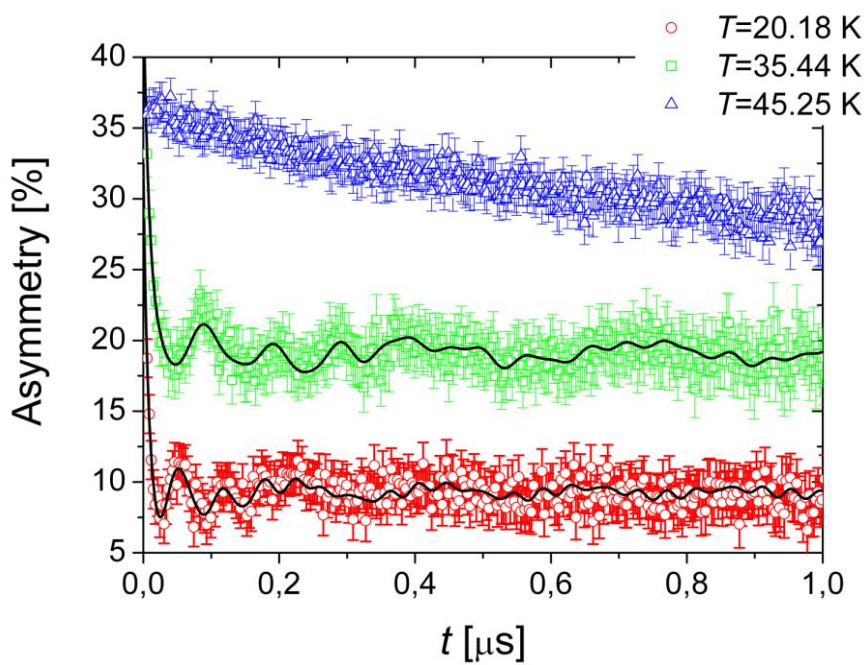

**Fig. 2.** Time dependence of the ZF asymmetry function detected in the vicinity of the transition point. The onset of spontaneous oscillations in the relaxation signal below 46 K reveals the transition to a magnetically ordered state. Solid lines show the fits to the sum of several damped oscillatory components.

Spontaneous oscillations observed in the time dependence of the asymmetry function depicted in Fig. 2, reflecting the muon spin precession in the quasistatic local magnetic field, provide unambiguous evidence of the long-range magnetic order in that temperature region. The frequency of the oscillations is equal to that of the precession. Three precession frequencies were observed in the measured asymmetry spectra implying several magnetically unique stopping sites in the material. The three data sets of local fields in Fig. 1 were simultaneously fitted to the phenomenological form

$$B(T) = B(0)\left[1 - \left(\frac{T}{T_c}\right)^\sigma\right]^\beta, \qquad (8)$$

where exponent $\sigma$ corresponds to the low-temperature properties governed by spin-wave excitations [19]. The best fit yielded $\sigma = 1.55(5)$, $\beta = 0.38(1)$, $T_c = 42.08(3)$ K, $B_1(0) = 399(12)$ G, $B_2(0) = 1312(20)$ G, and $B_3(0) = 2312(38)$ G. The value of the exponent $\beta$ falls very close to that corresponding to the three-dimensional (3D) Heisenberg model [20]. The parameter $\sigma$ is consistent with the value of 3/2 expected for ferromagnetic magnons, although compound **1** was demonstrated to be a ferrimagnet. However, for ferrimagnets in a bipartite lattice the elementary magnetic excitations are known to split into two branches [21-23], of which one in accordance with the Goldstone theorem starts off at zero (gapless acoustic magnons), while the energy in the other branch remains finite for all values of the wave vector (gapped optical magnons). The acoustic modes, exhibiting a quadratic (ferromagnetic) dispersion relation in the long-wavelength limit, determine the thermal behaviour of the total magnetization at low temperatures giving rise to the $T^{3/2}$ Bloch-like dependence of the relative reduction of the total magnetization above 0 K. Due to the Goldstone theorem one acoustic branch with ferromagnetic dispersion relation will be likewise present for the reported compound.

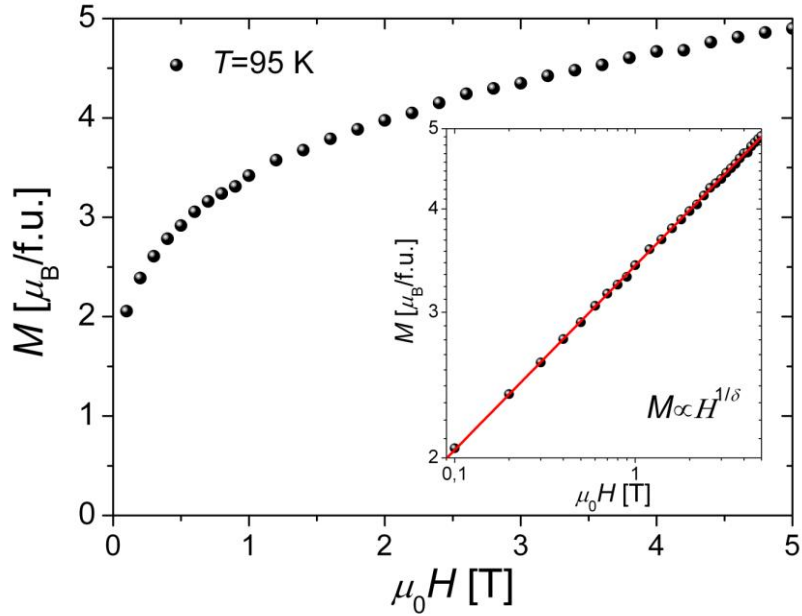

**Fig. 3.** Isothermal magnetization for **2** measured very close to the transition temperature $T_c$=95.25 K. Inset: Log-log plot of the $M$ vs $H$ dependence. The inverse of the slope yields the critical-point exponent $\delta = 4.49(1)$.

In contrast to $\beta$ critical-point exponent $\delta$ may be accessed by a magnetometric technique. It is namely sufficient to measure the field dependence of isothermal magnetization at the critical temperature $M(T_c, H)$. Then the low-field data should display the power-law behaviour. Figure 3 shows the field dependence of the isothermal magnetization of **2** detected very close to the transition temperature $T_c = 95.25$ K. Inset of Fig. 3 depicts the log-log plot of this dependence. One can see that for all values of magnetic field the experimental points collapse on a straight line. The slope of the line yields the inverse of the critical-point exponent $\delta = 1.49(1)$. This value is close to that expected for the 3D Heisenberg universality class (=4.783(3) [20]).

Another pair of exponents $\gamma'$ and $\gamma$ is related to the behaviour of the zero-field susceptibility near the critical point. Corresponding definitions read

$$\chi(\varepsilon, H = 0) \propto \begin{cases} (-\varepsilon)^{-\gamma'} & \text{for } T < T_c \ (\varepsilon \to 0^-), \\ \varepsilon^{-\gamma} & \text{for } T > T_c \ (\varepsilon \to 0^+), \end{cases} \qquad (9)$$

where we distinguish whether the critical point is being approached from above or from below. The direct way to access these critical-point exponents experimentally is standard magnetometry represented by either dc magnetization or ac susceptibility. While the applied magnetic field necessary for the dc measurements may significantly distort the signal in the critical region, the ac susceptibility measurements involving the sweeping fields of a much smaller magnitude seem to be better suited to the inspection of the critical behaviour. Before processing the experimental data to the form of the log-log plot, one important issue must be resolved, namely the determination of the position of the critical temperature $T_c$. The task is straightforward if the ac susceptibility signal is sufficiently sharp. If it is not the case one may resort to other experimental techniques, like the ZF muon spin rotation spectroscopy, which provide a precise pinpointing of the onset of magnetic order. Another possible approach is based on a statistical analysis of the ac signal. In this approach one fixes the value of $T_c$, next performs the log-log plot of $\chi_{ac}(\varepsilon)$ with this value, and then carries out the fitting of a straight line and calculates some measure of the goodness of the fit, like, e.g., estimated variance. This procedure should be repeated for $T_c$'s in some interval encompassing the ac susceptibility anomaly. The actual critical temperature is finally selected as that corresponding to the lowest value of the estimated variance. Such a procedure was performed for the ac susceptibility data of **3** above the transition temperature. Figure 4 shows the log-log plot of the ac susceptibility data with the best-fit line for $T_c = 32.75(25)$ K corresponding to the lowest value of the estimated variance.

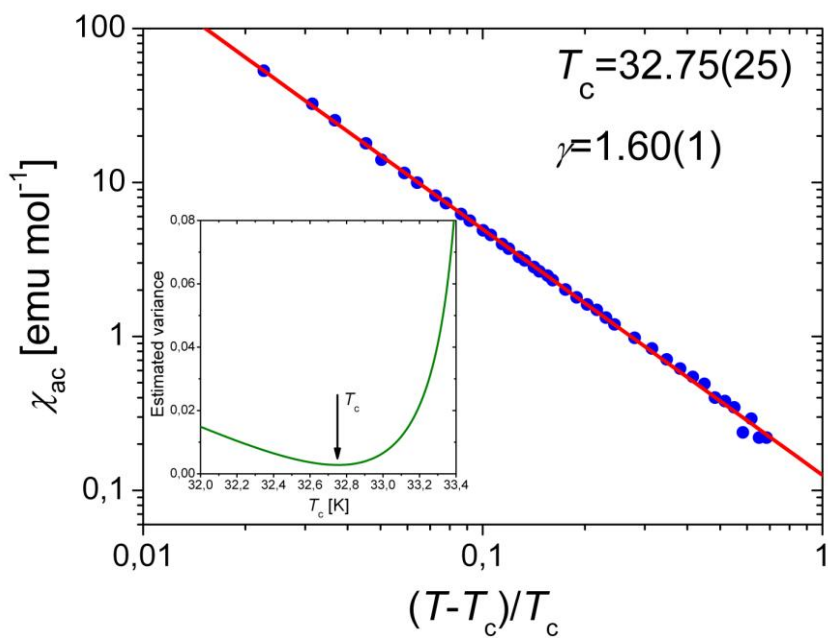

**Fig. 4.** Log-log plot of $\chi_{ac}$ vs $\varepsilon$ for the optimal value of $T_c$. The solid line represents the best linear fit to the experimantal data. Inset: The estimated variance of linear fits to the log-log data of $\chi_{ac}$ vs $\varepsilon$ for different choices of $T_c$.

Further two critical exponents $\alpha'$ and $\alpha$ are related to the asymptotic behaviour of the specific heat. They are defined by the following relations

$$C(\varepsilon, H=0) \propto \begin{cases} (-\varepsilon)^{-\alpha'} & \text{for } T < T_c \ (\varepsilon \to 0^-), \\ \varepsilon^{-\alpha} & \text{for } T > T_c \ (\varepsilon \to 0^+). \end{cases} \quad (10)$$

It is important to note that the exponents associated with the critical behaviour of heat capacity may be positive, negative or vanish. The vanishing $\alpha$ exponents correspond to a logarithmic singularity present, e.g., in the two-dimensional Ising model. For this reason, if one wants to determine the exponents, a more general definition encompassing the three different cases is recommended, i.e.

$$C(\varepsilon, H=0) = \frac{A}{\alpha}\left(|\varepsilon|^{-\alpha} - 1\right) \quad (11)$$

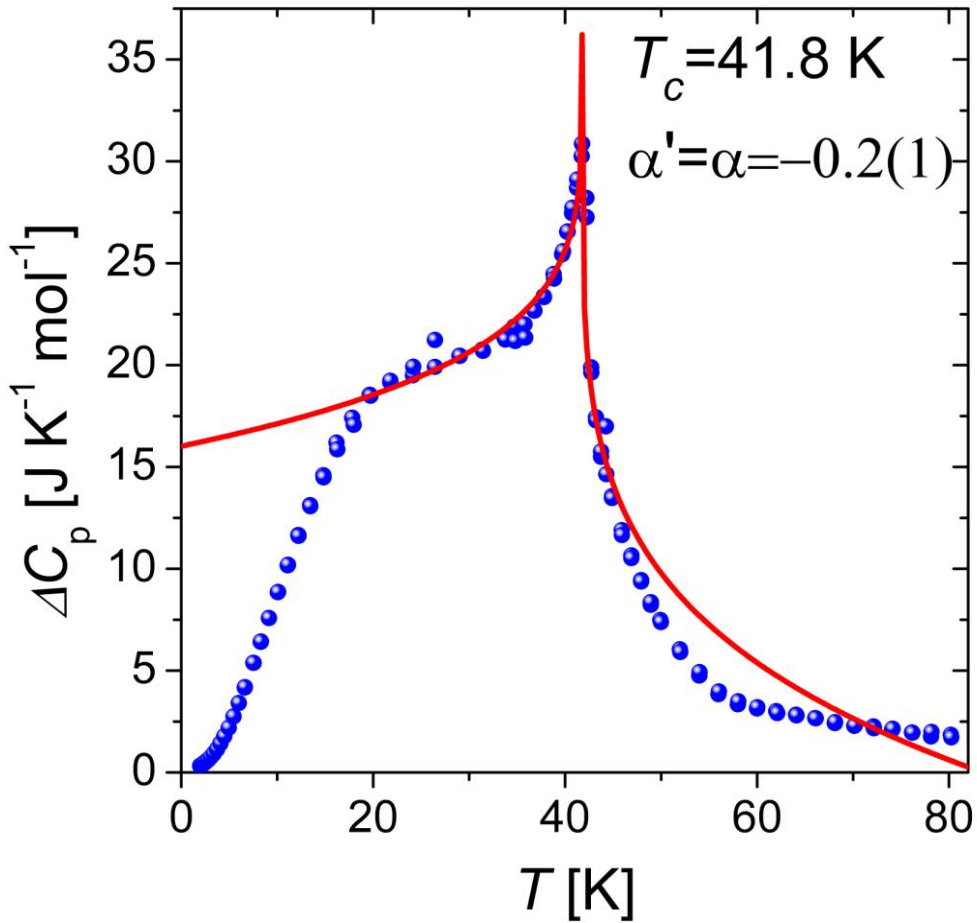

**Fig. 5.** Magnetic excess heat capacity of **1**. Solid lines represent the best fits to the scaling function. A negative value of critical exponent $\alpha$ is consistent with the 3D Heisenberg model.

The scaling analysis of the excess heat capacity $\Delta C_p$ of **1** was performed by fitting the function given in Eq. (11) independently above $T_c$ ($T < 50$ K) and below $T_c$ ($T > 20$ K). To account for the elevated heat capacity shoulder for $T < T_c$ an additive constant $B'$ was introduced. The best fits (solid lines in Fig. 5) yielded $A' = 4(1)$ J K$^{-1}$ mol$^{-1}$, $B' = 16(1)$ J K$^{-1}$ mol$^{-1}$, $\alpha' = -0.2(1)$, $A = 7.1(3)$ J K$^{-1}$ mol$^{-1}$, and $\alpha = -0.20(2)$, where primes refer to $T < T_c$. The result is consistent with the scaling prediction $\alpha' = \alpha$ (see Section 3). The negative exponents imply that $\Delta C_p$ is finite at $T_c$ with amplitudes equal to $B' - A'/\alpha' = 36(16)$ J K$^{-1}$ mol$^{-1}$ and $-A/\alpha = 35.5(6)$ J K$^{-1}$ mol$^{-1}$, respectively, which is in agreement with the continuous character of the second-order phase transition. The value of exponent $\alpha$ is close to that expected for the 3D Heisenberg model ($\alpha = -0.14$ [24]).

Let us complete this section with the definition of less common critical-point exponents which, nevertheless, will be demonstrated in Section 6 to emerge in the context of scaling properties of magnetocaloric effect. For a magnetic system, the spontaneous magnetization and the zero-field susceptibility are proportional, respectively, to the first and second derivatives of the Gibbs potential $G(T, H)$ with respect to the magnetic field $H$ (evaluated at $H = 0$). In 1963 Essam and Fisher suggested [13] that one might consider higher field derivatives of $G(T, H)$, and define a sequence of exponents $\Delta'_l$ called gap exponents by relations

$$\left. \frac{\partial^l G}{\partial H^l} \right|_{\varepsilon \to 0^-, H=0} \equiv G^{(l)}(\varepsilon \to 0^-, H = 0) \propto \varepsilon^{-\Delta'_l} G^{(l-1)}(\varepsilon \to 0^-, H = 0). \qquad (12)$$

Since $G^{(1)} \propto M \propto \varepsilon^\beta$ and the heat capacity $C_{H=0} \propto \partial^2 G/\partial T^2$ diverges as $\varepsilon^{-\alpha'}$, it follows that $G^{(0)} \propto \varepsilon^{2-\alpha'}$ and hence $\Delta'_1 = 2 - \alpha' - \beta$. Similarly, the fact that $G^{(2)} \propto \chi \propto \varepsilon^{-\gamma'}$ means that $\Delta'_2 = \beta + \gamma'$. In Section 3 certain symmetry arguments will be provided which imply the equality of $\Delta'_l$ for all $l$.

3. The static scaling hypothesis

The discussion of the critical behaviour would not be complete without the illuminating conjecture of the scaling hypothesis [25,26]. The static scaling hypothesis for thermodynamic functions is expressed in the form of a statement about one particular thermodynamic potential, generally chosen to be the Gibbs potential, $G(T,H) = G(\varepsilon,H)$, where $\varepsilon$ is the reduced temperature defined in Eq. (1). It asserts that asymptotically close to the critical point the singular part of $G(\varepsilon,H)$ is a generalized homogeneous function (GHF). Thus, the statement is equivalent to the requirement that asymptotically close to the critical point there exist two parameters $a_T$ and $a_H$ (called the temperature and field scaling powers), such that for all positive $\lambda$ the Gibbs potential $G(\varepsilon,H)$ obeys the functional equation

$$G(\lambda^{a_T}\varepsilon, \lambda^{a_H} H) = \lambda G(\varepsilon, H). \tag{13}$$

Let us stress that the scaling hypothesis does not specify the parameters $a_T$ and $a_H$, which corresponds to the fact that the homogeneous function or scaling hypothesis does not determine the values of critical-point exponents. The fact that the Legendre transforms of GHF's are also GHF's implies that all thermodynamic potentials share this scaling property. Moreover, derivatives of GHF's are also GHF's and since every thermodynamic function is expressible as some derivative of some thermodynamic potential, it follows that the singular part of every thermodynamic function is asymptotically a GHF.

The scaling parameters $a_T$ and $a_H$ can be related to the various critical-point exponents. Differentiating both sides of Eq. (13) with respect to $H$ and using the relation $M(\varepsilon,H) = -(\partial G(\varepsilon,H)/\partial H)_T$ one obtains

$$\lambda^{a_H} M(\lambda^{a_T}\varepsilon, \lambda^{a_H} H) = \lambda M(\varepsilon, H). \tag{14}$$

There are two critical-point exponents associated with the behaviour of the magnetization near the critical point. Consider first the case where $H = 0$ and $\varepsilon \to 0^-$. Then Eq. (14) implies

$$M(\varepsilon,0) = \lambda^{a_H - 1} M(\lambda^{a_T},0). \tag{15}$$

Since Eq. (15) is valid for all values of positive number $\lambda$, it must certainly hold for the particular choice $\lambda = (-1/\varepsilon)^{1/a_T}$. Thus one arrives at

$$M(\varepsilon,0) = (-\varepsilon)^{\frac{1-a_H}{a_T}} M(-1,0). \tag{16}$$

By comparing Eq. (16) with the definition of exponent $\beta$ in Eq. (6) one obtains

$$\beta = \frac{1-a_H}{a_T}. \tag{17}$$

The other magnetization related exponent $\delta$, see the definition in Eq. (7), can also be expressed in terms of the scaling parameters by setting $\varepsilon = 0$ in Eq. (14) and letting $H \to 0$,

$$M(0,H) = \lambda^{a_H -1} M(0, \lambda^{a_H} H). \tag{18}$$

Now, choosing $\lambda = H^{-1/a_H}$, Eq. (18) becomes

$$M(0,H) = H^{\frac{1-a_H}{a_H}} M(0,1). \tag{19}$$

Hence, one arrives at the relation

$$\delta = \frac{a_H}{1-a_H}. \tag{20}$$

One can obtain additional exponents by forming further derivatives of the Gibbs potential. On differentiating twice with respect to $H$, one obtains

$$\lambda^{2a_H} \chi(\lambda^{a_T} \varepsilon, \lambda^{a_H} H) = \lambda \chi(\varepsilon, H), \tag{21}$$

where $\chi$ is the isothermal susceptibility. If one considers $H = 0$ and chooses $\lambda = (-\varepsilon)^{-1/a_T}$, Eq. (21) becomes

$$\chi(\varepsilon,0) = (-\varepsilon)^{-\frac{2a_H-1}{a_T}} \chi(-1,0). \tag{22}$$

Comparison with the definition in Eq. (9) for the limit $\varepsilon \to 0^-$ yields

$$\gamma' = \frac{2a_H - 1}{a_T}. \tag{23}$$

Likewise, setting $H = 0$ and choosing $\lambda = \varepsilon^{-1/a_T}$ in Eq. (21) one gets

$$\chi(\varepsilon,0) = \varepsilon^{-\frac{2a_H-1}{a_T}} \chi(1,0), \tag{24}$$

whence, by comparison with the definition in Eq. (9) for the limit $\varepsilon \to 0^+$ one obtains

$$\gamma = \frac{2a_H - 1}{a_T}. \tag{25}$$

Combining Eqs. (23) and (25) we arrive at one of the hallmarks of static scaling hypothesis, i.e. the equality of primed and unprimed critical-point exponents.

Differentiating Eq. (13) with respect to temperature and using the relation $C = -T(\partial^2 G/\partial T^2)_H$ one arrives at the equation

$$\lambda^{2a_T} C(\lambda^{a_T} \varepsilon, \lambda^{a_H} H) = \lambda C(\varepsilon, H). \tag{26}$$

On setting $H=0$ and $\lambda = (\mp\varepsilon)^{1/a_T}$ and comparing with the definition in Eq. (10) one obtains

$$\alpha' = \alpha = \frac{2a_T - 1}{a_T}. \tag{27}$$

Since each critical-point exponent, as exemplified above, is directly expressible in terms of two unknown scaling parameters $a_T$ and $a_H$, it follows that one can eliminate these two scaling powers from the expressions for three different exponents, and thereby obtain a family of equalities called scaling laws. Performing such an elimination in Eqs. (17), (23), and (27), we find

$$\alpha' + 2\beta + \gamma' = 2, \tag{28}$$

which is the Rushbrooke scaling law expressed originally as an inequality $\alpha' + 2\beta + \gamma' \geq 2$. Furthermore, eliminating $a_T$ and $a_H$ from Eqs. (17), (20), and (27), we arrive at the Griffiths scaling law

$$\alpha' + \beta(\delta + 1) = 2. \tag{29}$$

Similarly, Eqs. (17), (20), and (23) give the Widom equality

$$\gamma' = \beta(\delta - 1). \tag{30}$$

In general, it suffices to determine two exponents since these will fix the scaling powers $a_T$ and $a_H$, which in turn may be used to obtain exponents for any thermodynamic function. Furthermore, the static scaling hypothesis can be used to demonstrate that the gap exponents $\Delta'_l$ are all equal. If we differentiate both sides of Eq. (13) $l$ times with respect to $H$, we obtain

$$G^{(l)}(\varepsilon, H) = \lambda^{la_H - 1} G^{(l)}(\lambda^{a_T}\varepsilon, \lambda^{a_H} H). \tag{31}$$

Hence it follows that

$$\frac{G^{(l)}(\varepsilon, H)}{G^{(l-1)}(\varepsilon, H)} = \lambda^{a_H} \frac{G^{(l)}(\lambda^{a_T}\varepsilon, \lambda^{a_H} H)}{G^{(l-1)}(\lambda^{a_T}\varepsilon, \lambda^{a_H} H)}. \tag{32}$$

Now, the definition of the gap exponents in Eq. (12) implies that when $H=0$ the left hand side of Eq. (32) varies as $\varepsilon^{-\Delta'_l}$. Therefore on choosing $\lambda = \varepsilon^{-1/a_T}$ we obtain $\Delta'_l = \Delta$, independent of order $l$, where

$$\Delta = \frac{a_H}{a_T} = \beta\delta = \beta + \gamma'. \tag{33}$$

In addition to predicting the relations among the critical-point exponents, the scaling hypothesis makes specific predictions concerning the form of the magnetic equation of state, i.e. the relation among the variables $M$, $H$, and $T$. Setting $\lambda = |\varepsilon|^{-1/a_T}$ in Eq. (14) we obtain

$$M(\varepsilon, H) = |\varepsilon|^{\frac{1-a_H}{a_T}} M\left(\frac{\varepsilon}{|\varepsilon|}, \frac{H}{|\varepsilon|^{a_H/a_T}}\right). \tag{34}$$

Using Eqs. (17) and (33) to eliminate scaling parameters in favour of the critical-point exponents Eq. (34) becomes

$$\frac{M(\varepsilon, H)}{|\varepsilon|^\beta} = M\left(\frac{\varepsilon}{|\varepsilon|}, \frac{H}{|\varepsilon|^\Delta}\right). \tag{35}$$

Let us note that the function on the right-hand side of Eq. (35) is a function of the scaled magnetic field $h = |\varepsilon|^{-\Delta} H(\varepsilon, M)$ and the sign of the reduced temperature $\varepsilon$ ($\varepsilon/|\varepsilon| = \pm 1$). We may thus define

$$m_\pm(h) = M(\pm 1, h), \tag{36}$$

and Eq. (35) may be written

$$m = m_\pm(h), \tag{37}$$

where $m = |\varepsilon|^{-\beta} M(\varepsilon, H)$ is the scaled magnetization. Eq. (37) predicts that if we rescale magnetization $M$ by dividing by $|\varepsilon|^\beta$ and rescale magnetic field $H$ by dividing by $|\varepsilon|^\Delta$, then plots of $m$ vs $h$ should collapse on two universal branches, one corresponding to temperatures below and the other to temperatures above the critical temperature $T_c$, in contrast to the case of plots $M$ vs $H$, where the data fall on distinct isotherms. An example of this scaling equation is provided in Fig. 6.

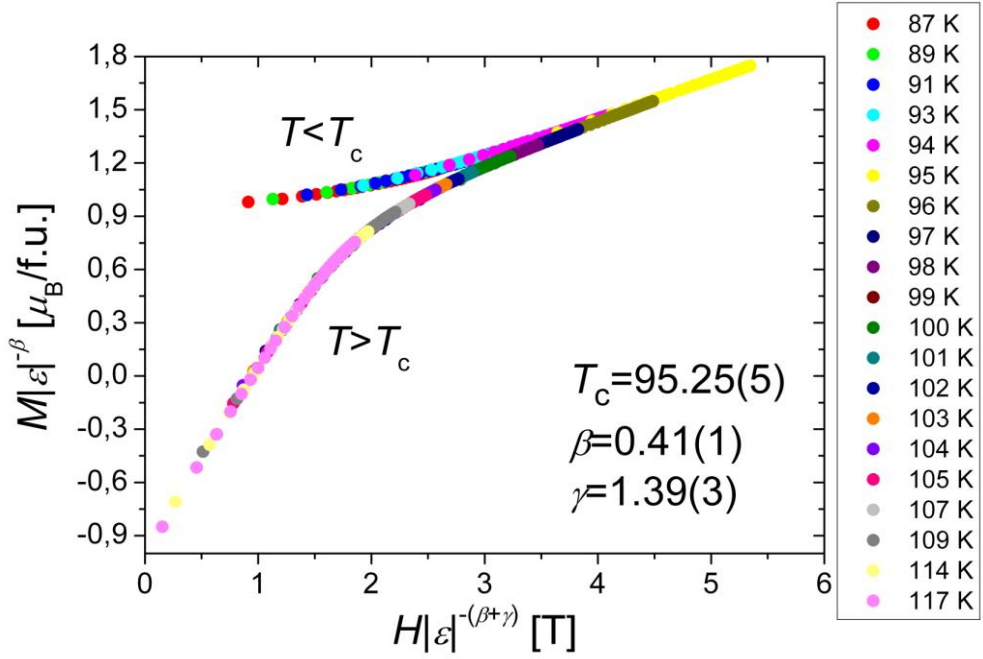

**Fig. 6.** Logarithmic scaling plot of $M|\varepsilon|^{-\beta}$ vs $H|\varepsilon|^{-(\beta+\gamma)}$ for **2** evidences the validity of the critical exponents and the $T_c$ value.

Another form of the scaling equation of state, leading to a single universal curve, can be obtained, if one chooses $\lambda = H^{-1/a_H}$ in Eq. (14). Then it becomes

$$\frac{M(\varepsilon,H)}{H^{(1-a_H)/a_H}} = M\left(\frac{\varepsilon}{H^{a_T/a_H}},1\right). \qquad (38)$$

Eliminating the scaling powers $a_T$ and $a_H$ with the use of Eqs. (20) and (33) we arrive at

$$\frac{M(\varepsilon,H)}{H^{1/\delta}} = M\left(\frac{\varepsilon}{H^{1/\Delta}},1\right). \qquad (39)$$

Defining the scaled magnetization by $\tilde{m} = H^{-1/\delta} M(\varepsilon,H)$ and the scaled temperature by $\tilde{\varepsilon} = H^{-1/\Delta}\varepsilon$, one may write the ensuing scaling equation of state in the form

$$\tilde{m} = f(\tilde{\varepsilon}), \qquad (40)$$

where $f(\tilde{\varepsilon}) \equiv M(\tilde{\varepsilon},1)$ is the scaling function. Let us note that for this form of the equation of state there is no distinction between the temperature ranges below and above the critical temperature $T_c$. Only one scaling function is present in the equation implying that if one plots the scaled magnetization $\tilde{m}$ against the scaled temperature $\tilde{\varepsilon}$ the entire family of $M(T,H)$ curves will collapse onto a single universal curve. This form of the equation of state is depicted in Fig. 7 for the same compound **2**.

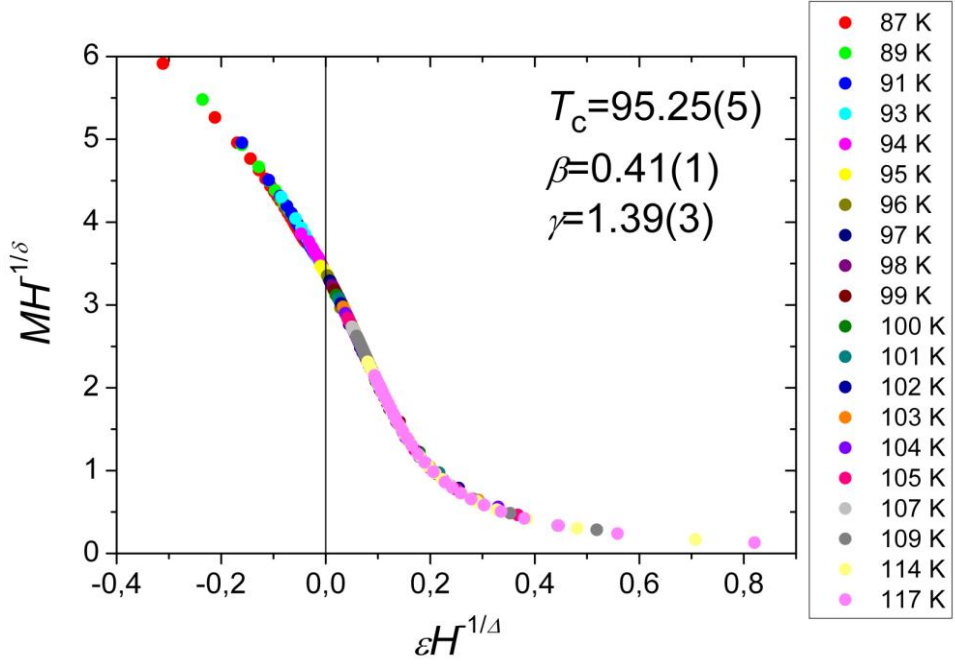

**Fig. 7.** Scaling plot of $MH^{-1/\delta}$ vs $\varepsilon H^{-1/\Delta}$ for **2**.

4. The Kouvel-Fisher approach

Performing log-log plots of the thermodynamic functions near the critical point is the most straightforward method of determining the associated critical exponents. Another method was proposed by Kouvel and Fisher [27]. It consists of an iterative procedure in which the Arrott-Noakes plot, i.e. the plot of $M^{2.5}$ versus $(H/M)^{0.75}$, is constructed [28,29]. From this plot the values for spontaneous magnetization $M_0(T)$ are computed from the intercepts of various isothermal magnetization vs field curves on the ordinate of the plot for temperatures below the transition temperature $T_c$. The intercept on the abscissa allows one to calculate the initial susceptibility $\chi_0(T)$ for temperatures above $T_c$. Once the $M_0(T)$ and $\chi_0(T)$ curves have been constructed, two additional quantities $X(T)$ and $Y(T)$ are determined

$$X(T) = \chi_0^{-1}\left(\frac{d\chi_0^{-1}}{dT}\right)^{-1} = \frac{T-T_c}{\gamma}, \tag{41}$$

$$Y(T) = M_0\left(\frac{dM_0}{dT}\right)^{-1} = \frac{T-T_c}{\beta}. \tag{42}$$

In the critical region both $X(T)$ and $Y(T)$ should be linear with slopes which give the values of the critical exponents and intercepts with the temperature axis that correspond to the critical temperature. The values of critical exponents are refined using an iterative method; using the critical exponents found from Eqs. (41) and (42) a generalized Arrott-Noakes plot ($M^{1/\beta}$ vs $(H/M)^{1/\gamma}$) is constructed, see Fig. 8, and used to claculate new $M_0(T)$ and $\chi_0(T)$ curves, which are subsequently input into Eqs. (41) and (42), resulting in a new set of values for $\beta$ and $\gamma$. The procedure finishes when the desired convergence of the parameters is achieved. Figure 9 shows the final iteration step for **2**. The values extracted from this plot are $\beta = 0.41(1)$, $\gamma = 1.39(3)$, and $T_c = 95.25(5)$ K.

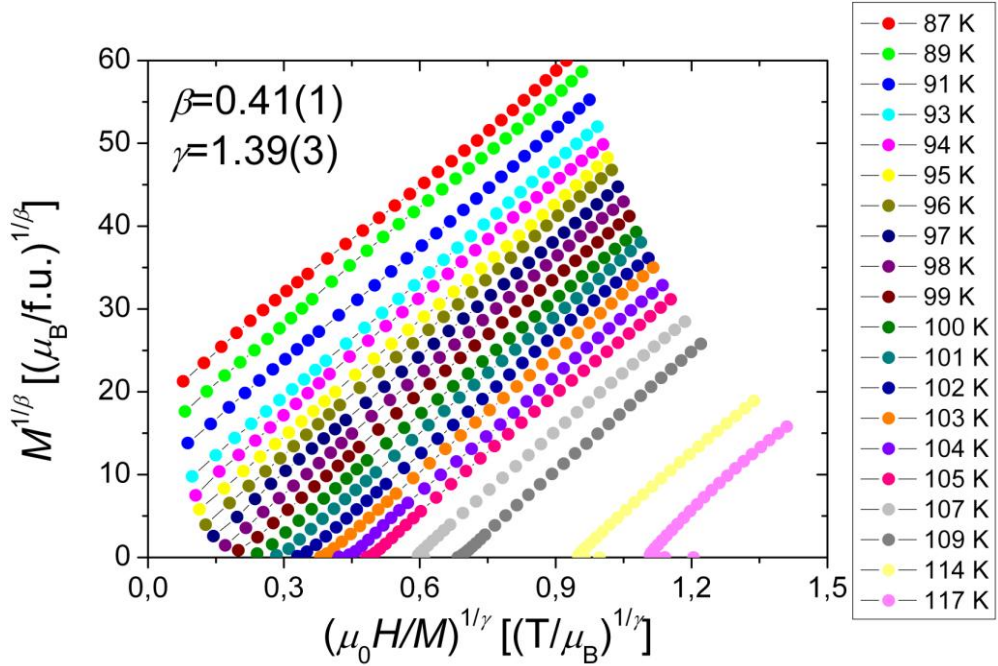

**Fig. 8.** Generalized Arrott-Noakes plot for **2** using the exponents extracted from the Kouvel-Fisher analysis. The different colors correspond to the data measured at temperatures ranging from 87 to 117 K.

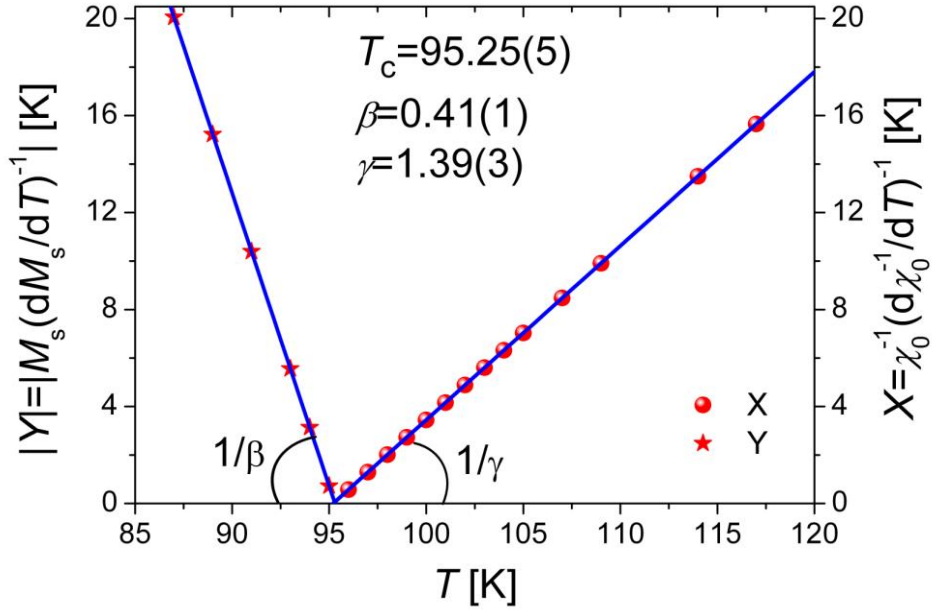

**Fig. 9.** Determination of the critical exponents and the critical temperature for **2** using the Kouvel-Fisher method.

The reliability of thus obtained exponents and the critical temperature can be ascertained by checking the scaling of the magnetization curves using equation of states given in Eqs. (37) and (40), see Fig. 6 and 7. Application of this method requires the knowledge of field and temperature dependence of the magnetization $M(T,H)$ near the critical point. This can be achieved by measuring magnetization isotherms in a certain interval of magnetic field and for an array of temperatures surrounding the critical temperature $T_c$. Apart from critical exponents the method yields simultaneously the position of the transition temperature $T_c$.

5. The generalized Curie-Weiss law

In 1983 Souletie and Tholence [30] pointed out that the temperature dependence of the paramagnetic susceptibility $\chi$ of crystalline nickel may be very well represented by a power law over a wide temperature range above the Curie temperature $T_c$. This was a great surprise, because usually it is assumed that a power law is only valid in the small critical region very close to $T_c$, whereas the temperature dependence outside the universality range is expected to be rather complicated. Fähnle and Souletie and independently Arrott have shown that [31-34] the Padé approximants and the high-temperature series expansions, respectively, for the susceptibility of some localized spin models indeed are very well approximated by a power law over the whole temperature range. The only prerequisite to have an extended domain where scaling ideas are practical is that the nonlinear scaling variable

$$\varepsilon' = \frac{\varepsilon}{1+\varepsilon} = \frac{T-T_c}{T} \qquad (43)$$

be used instead of the linear one $\varepsilon = (T-T_c)/T_c$. Then the scaling law for the susceptibility may be summarized by a formula, called the generalized Curie-Weiss law by Arrott, which is most generally written in the form

$$\chi T = C\left(1 - \frac{T_c}{T}\right)^{-\gamma}, \qquad (44)$$

where $C$ is the Curie constant. For example, the susceptibility data for crystalline nickel were successfully fitted to Eq. (44) with $\gamma = 1.31$ up to $T = 3T_c$. By linearizing Eq. (44) we arrive at the mean-field limit

$$\chi^{-1}(T \to \infty) \propto (T - T_c^{MF}) \qquad (45)$$

with a nontrivial prediction $T_c^{MF} = \gamma T_c$. Eq. (44) can be differentiated to obtain an equivalent expression

$$\frac{d \ln T}{d \ln(\chi T)} = -\frac{T-T_c}{\gamma T_c}, \qquad (46)$$

which has only two parameters. A linear regime in the representation of $d\ln T / d\ln(\chi T)$ vs $T$ means that the scaling is valid in the corresponding domain with the parameters $\gamma^{-1}$ and $T_c$, which are deduced simultaneously at the intercepts of the line with $T=0$ and $d\ln T / d\ln(\chi T) = 0$ axes, respectively. This form of scaling analysis turned out to be a very sensitive method giving unambiguously values for $\gamma$ and $T_c$ [35-37].

To have a precise insight into the character of the ordering process in **3** such an analysis of the static critical scaling has been performed. The magnetization data of a single crystal sample detected in the field of 2 kOe were appropriately rescaled to yield the dc magnetic susceptibility. Figure 10 shows the plot of $d\ln T / d\ln(\chi T)$ against temperature.

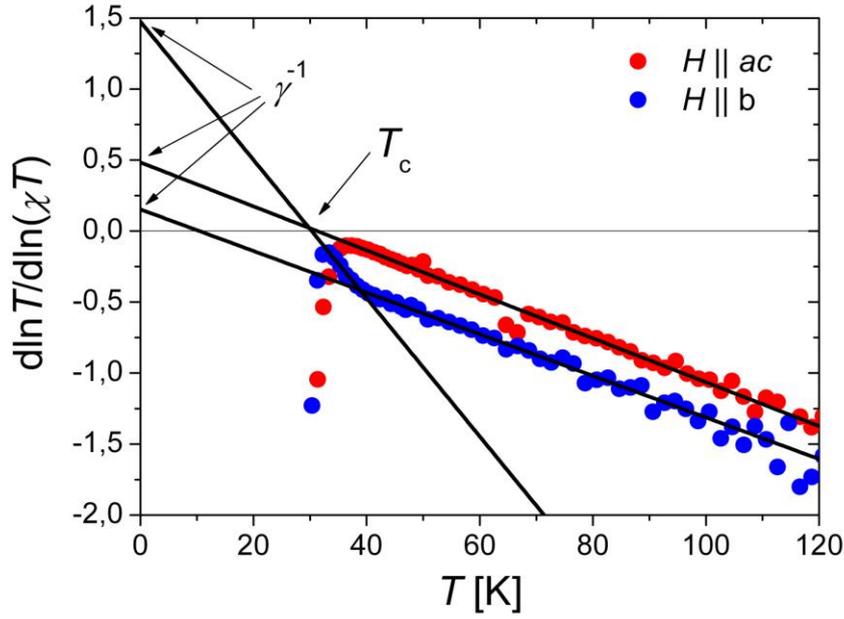

**Fig. 10.** Critical scaling analysis for **3**. The $\mathrm{d}\ln T/\mathrm{d}\ln(\chi T)$ vs $T$ plot for the direction parallel to the *ac* crystallographic plane and that perpendicular to that plane. In the latter case a crossover between two ordering regimes is apparent.

It is apparent that the data tend to align in the pretransitional high-temperature region. The linear fits revealed the values of the corresponding transition temperatures through the intercepts with the abscissa axis, and the critical exponents $\gamma$ through the inverses of the intercepts with the ordinate axis. In the direction parallel to the *ac* crystallographic plane the system undergoes a one-step transition at $T_{c\parallel} = 31.2 \pm 2.3\,\mathrm{K}$ displaying a rather high value of $\gamma_\parallel = 2.1 \pm 0.3$. This value fails to agree with the 3D ferromagnetic ordering processes ($\gamma \approx 1.24$ for the Ising model, 1.32 for the XY model, and 1.38 for the Heisenberg model). On the other hand, it is consistent with the values obtained in the numerical simulations of the 2D classical XY model ($\gamma = 1.82$) [38] or, more pertinent to the case, of the 2D classical XXZ model ($\gamma = 2.17 \pm 0.05$) [39]. The behaviour in the direction parallel to the *b* crystallographic axis reveals a crossover at $\approx 38.8\,\mathrm{K}$ from a region with an exceptionally high value of the gamma exponent $6.7 \pm 1.8$ and the fictitious transition temperature at $10.4 \pm 1.5\,\mathrm{K}$ to the state characterized by a low value of $\gamma_\perp = 0.67 \pm 0.04$ and the transition temperature $T_{c\perp} = 30.3 \pm 3.4\,\mathrm{K}$. The $T_c$'s for both directions have close values which points to the fact that the transition in the *ac* plane triggers that in the direction perpendicular to that plane. The unusually high value of $\gamma$ in the precrossover region suggest the scaling behaviour close to the exponential one, which is well known to be characteristic of 2D isotropic Heisenberg model with no transition occurring at finite temperature.

The size of the critical region may be defined as the range in which $\chi(T)$ is reasonably well described by the simple power law $\chi \propto \varepsilon^{-\gamma}$. As far as experimental work is concerned an accepted estimate of this regime is obtained using the effective exponent $\gamma^*(T)$ introduced by Kouvel and Fisher [27]

$$\gamma^*(T) = -(T - T_c)\frac{d \ln \chi}{dT}. \qquad (47)$$

By construction, the quantity $\gamma^*$ approaches the asymptotic critical value $\gamma$ for $T \to T_c$ and the mean-field value of $\gamma = 1$ for $T \gg T_c$. The simple power law $\chi \propto \varepsilon^{-\gamma}$ is an accurate representation as long as the deviations $\gamma^*(T) - \gamma$ are small. Reanalyzing the data for crystalline nickel Kouvel and Fisher found that $\gamma^*(T)$ deviates from the value 1.35 for $T = T_c$ to $\gamma^* = 1.25$ for $T = 1.1T_c$. This analysis therefore confirmed the general belief that the crossover to universality occurs in a temperature range $\varepsilon < 0.1$. For a generalized Curie-Weiss law a different definition of the effective exponent is required. It is given by the formula

$$\gamma^{**}(T) = \gamma^*(T)\frac{T}{T_c} - \left(\frac{T}{T_c} - 1\right) = -\left(\frac{T}{T_c} - 1\right)\frac{d \ln(\chi T)}{d \ln T}. \qquad (48)$$

Inserting the modified power law given in Eq. (44) into Eq. (48) one obtains $\gamma^{**} = \gamma$. In the mean-field limit which is naturally approached in all systems when $T \to \infty$, $\gamma^{**}(\infty)$ would be some number not necessarily equal to 1 (which would correspond to the mean-field system), with the mean-field result $\chi \propto (T - T_c^{MF})^{-1}$ becoming approximately valid and $T_c^{MF} = \gamma^{**}(\infty)T_c \neq T_c$. Figure 11 shows the temperature dependence of the effective exponent $\gamma^{**}$ as deduced in two crystallographic directions for **3**. For the data detected in the direction parallel to the *ac* crystallographic plane (red circles) it is particularly clear that the critical region extends as far as $\approx 3T_c$. The apparent stepwise change of the effective gamma exponent detected in the direction parallel to the *b* crystallographic axis is related with the 2D-3D crossover mentioned above.

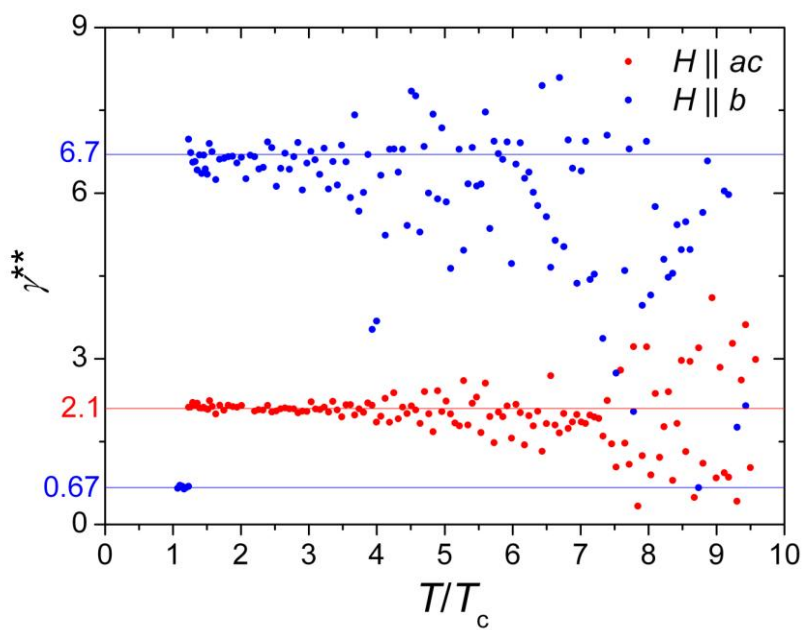

**Fig. 11.** Temperature dependence of the effective exponent $\gamma^{**}$ for **3** in two independent crystallographic directions

6. Scaling behavior of magnetocaloric effect

The magnetocaloric effect (MCE), i.e. the temperature change of a system when it is magnetized/demagnetized, is an intrinsic property of magnetic materials. Refrigerators based on MCE are expected to have an enhanced efficiency and to be more environmentally friendly than those based on gas compression-expansion, hence, MCE attracts increasing attention of researchers. Current trends in materials science related to this field go through the enhancement of materials performance (mostly associated with giant magnetocaloric effect GMCE [40,41]) and cost reduction (by replacing rare earths by transition metal based alloys [42]). With a view to gaining further clues of how to improve the performance of refrigerant materials, the field dependence of this effect is also being studied intensively, both experimentally [43,44] and theoretically [45-49]. MCE has also been investigated for molecular magnets, especially for single molecule magnets (SMMs), where a substantial entropic effect was anticipated due to their large grand-state spin value [50-53]. Apart from that, first studies of MCE driven by the transition to a long-range magnetically ordered phase dealt with Prussian blue analogues [54-56]. Recent examples refer to an interesting instance of a molecular sponge changing reversibly the ordering temperature and the coercive field upon hydration/dehydration [57], and a couple of molecular magnets, bimetallic octacyanoniobates with manganese and nickel, isomorphous with **4** [58].

Two main quantities characterizing MCE is the isothermal magnetic entropy change $\Delta S_M(T,\Delta H)$ and the adiabatic temperature change $\Delta T_{ad}(S,\Delta H)$ due to the change of the magnetic field $\Delta H = H_f - H_i$. From now on let us set $H_f = 0$ and $H_i = H$, which corresponds to the experimentally plausible procedure of switching off the external magnetic field. Then $\Delta H = -H$ and $\Delta S_M(T,-H)$ can be obtained either from temperature dependence of the heat capacity of a material by carrying out an appropriate subtraction

$$\Delta S_M(T,-H) = \int_0^T \frac{C_p(T',0) - C_p(T',H)}{T'} dT', \qquad (49)$$

where $C_p$ denotes the molar heat capacity of a sample at constant pressure, or from the processing of the temperature and field dependent magnetization curves $M(T,H)$, using the Maxwell relation

$$\Delta S_M(T,-H) = \int_H^0 \left(\frac{\partial M}{\partial T}\right)_{H'} dH'. \qquad (50)$$

Let us note that the choice of initial and final values of the magnetic field assures that the the effect is positive in terms of the magnetic entropy change $\Delta S_M$, which is implied by Eqs. (49) and (50). Typically, the $\Delta S_M$ vs $T$ curve displays a peak near the transition temperature $T_c$. The height of this peak depends on the magnitude of the magnetic field change $|\Delta H|$, and increases with the increase in $|\Delta H|$. It has been found that there exists a phenomenological universal curve for the field dependence of $\Delta S_M$ [47]. Its construction is based on the assumption that, if such a universal curve exists, equivalent points of the different $\Delta S_M(T,-H)$ curves measured up to different maximum applied fields should collapse onto the same point of the universal curve. The selection of the equivalent points of the experimental curves is based on the choice of the peak entropy change $\Delta S_M^{peak}$ as a reference point. It is assumed that all the points that are at the same level with respect to $\Delta S_M^{peak}$ should be in an equivalent state. In this way two different reference points are found for each curve, one below $T_c$ and the other above. After normalizing the curves with respect to their peaks, the test for the existence of the universal curve would be to impose a scaling law for the temperature axis, which makes equivalent points collapse and check if the remaining parts of the curves also collapse. The temperature axis is rescaled in a different way below and above

$T_c$, just by imposing that the position of the two reference points of each curve correspond to $\theta = \pm 1$, where

$$\theta = \begin{cases} -\dfrac{T-T_c}{T_{r1}-T_c} & T \leq T_c, \\ \dfrac{T-T_c}{T_{r2}-T_c} & T > T_c, \end{cases} \quad (51)$$

and $T_{r1}$ and $T_{r2}$ are the temperatures of the reference points of each curve defined by the equations

$$\frac{\Delta S_M(T_{r1})}{\Delta S_M^{peak}} = \frac{\Delta S_M(T_{r2})}{\Delta S_M^{peak}} = h, \quad (52)$$

where $h<1$ is the height of the equivalent states in the $\Delta S_M / \Delta S_M^{peak}$ curves, which is chosen in such a way that the curves to be overlapped have experimental values above that reference entropy change for temperatures below and above $T_c$. Figure 12(a) shows the temperature dependence of the magnetic entropy change inferred from the $M(T,H)$ data of **1**. In Fig. 12(b) the corresponding phenomenological universal curve for the normalized magnetic entropy change $\Delta S_M / \Delta S_M^{peak}$ is depicted. The apparent scaling behaviour confirms the transition of the second order in this compound.

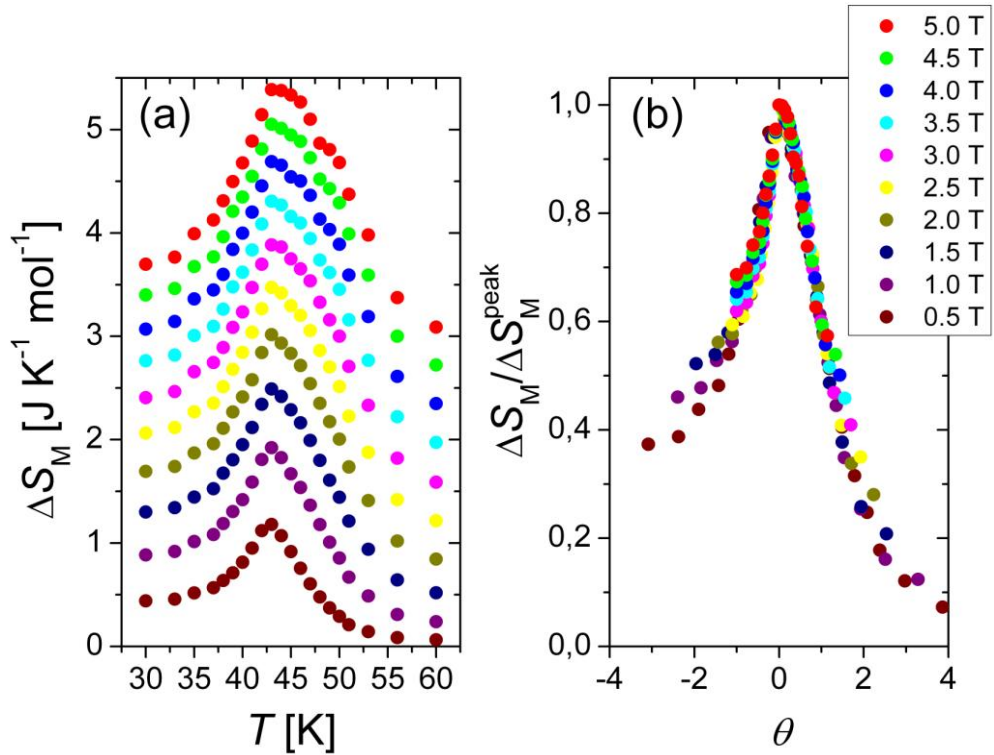

**Fig. 12.** (a) Temperature dependence of the magnetic entropy change $\Delta S_M$ for **1**. (b) The corresponding universal curve for the normalized magnetic entropy change.

The existence of the universal curve for second-order phase transitions has been already justified theoretically. The justification is based on the assumption that different physical magnitudes (such as magnetization) scale in the vicinity of a second-order transition and for magnetic systems the scaling equation is given in Eq. (37), where the plus (minus) sign corresponds to $\varepsilon > 0$ ($\varepsilon < 0$). Using Eq. (37) on some algebra Eq. (50), defining the magnetic entropy change due to the removal of a magnetic field $H$, can be transformed to the following form [47]

$$\frac{\Delta S_M}{a_M} = \mp |\varepsilon|^{1-\alpha} \int_0^{H/|\varepsilon|^\Delta} dx \left( \beta m_\pm(x) - \Delta x m'_\pm(x) \right)$$

$$= |\varepsilon|^{1-\alpha} \tilde{s}\left(\frac{\varepsilon}{H^{1/\Delta}}\right) = H^{\frac{1-\alpha}{\Delta}} s\left(\frac{\varepsilon}{H^{1/\Delta}}\right), \quad (53)$$

where $a_M = T_c^{-1} A^{\delta+1} B$, and $A$ and $B$ are critical amplitudes defined by relations $M = A(-\varepsilon)^\beta$ and $H = BM^\delta$, respectively. Note that $s(x) = |x|^{1-\alpha} \tilde{s}(x)$, and Eqs. (29) and (33) imply the identity $1 - \alpha = \beta + \Delta - 1$. Eq. (53) shows that if the reduced temperature $\varepsilon$ is rescaled by a factor proportional to $H^{1/\Delta}$, and the magnetic entropy change $\Delta S_M$ by $a_M H^{(1-\alpha)/\Delta}$, the experimental data should collapse onto the same curve. In this way the universal curve can be constructed analytically if only the critical exponents and the Curie temperature of a material are known. However, when characterizing the magnetocaloric response of a new material the critical exponents are not known a priori. Therefore, the phenomenological approach to constructing the universal curve may often come in useful. Similarly, the exponent $n$ controlling the field dependence of the magnetic entropy change, i.e.

$$\Delta S_M \propto H^n, \quad (54)$$

has the following scaling behavior

$$n = \frac{\partial \ln |\Delta S_M|}{\partial \ln H} = \frac{1-\alpha}{\Delta} - \frac{1}{\Delta} \frac{d \ln |s(x)|}{d \ln x}\bigg|_{x=\varepsilon/H^{1/\Delta}}. \quad (55)$$

Consequently, the values of $n$ should also collapse onto a universal curve when plotted against the same rescaled temperature axis for which the normalized values of $\Delta S_M$ collapse onto the same universal curve. Experimental evidence of this collapse of $n$ has been given for soft magnetic amorphous alloys [59,60]. Eq. (55) proves also that the field dependence of the magnetic entropy change at $T = T_c$,

$$\Delta S_{\mathrm{M}}\big|_{T=T_{\mathrm{c}}} \propto H^{\frac{1-\alpha}{\Delta}} = H^{1+\frac{1}{\delta}\left(1-\frac{1}{\beta}\right)}, \tag{56}$$

which was first derived from the Arrott-Noakes equation of state [47], is valid for any magnetic system following a scaling equation of state. Let us note that there is another temperature making the second term in Eq. (55) vanish, i.e. $T = T_{\mathrm{peak}}$, because in that case $\Delta S_{\mathrm{M}}$ has a peak implying $\mathrm{d}s(x)/\mathrm{d}x = 0$. Therefore, the field dependences of the magnetic entropy change at the critical temperature and at the temperature of the peak of the $\Delta S_{\mathrm{M}}$ curve are exactly the same. In between those temperatures exponent $n$ reaches its minimum value. In general the temperature $T_{\mathrm{peak}}$ at which the magnetic entropy change $\Delta S_{\mathrm{M}}$ attains a maximum for a given magnetic field change may not coincide with the transition temperature $T_{\mathrm{c}}$. This is demonstrated in Fig. 13, where the temperature dependence of the magnetic entropy change $\Delta S_{\mathrm{M}}$ inferred from the heat capacity measurements for **4** is shown.

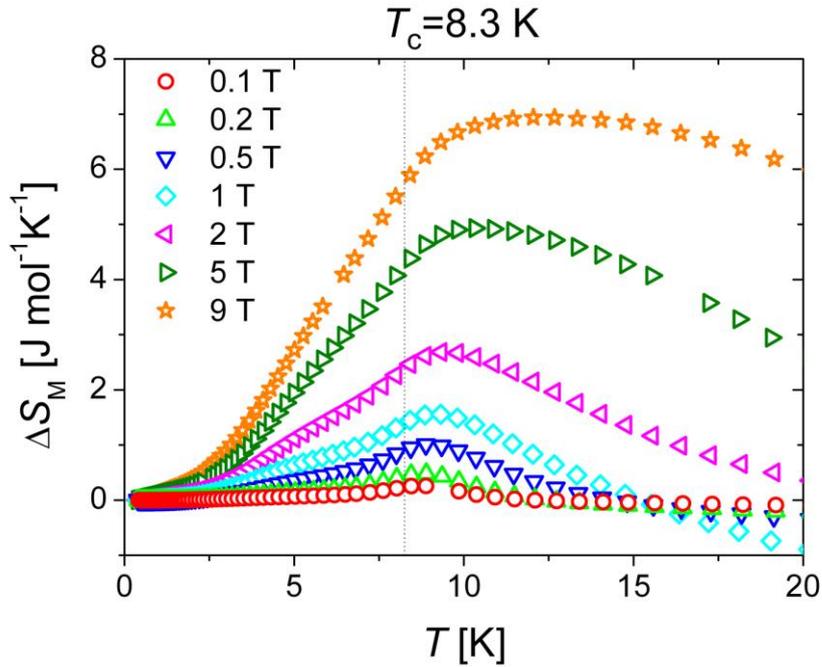

**Fig. 13.** Temperature dependence of the magnetic entropy change of compound **4** as inferred from the heat capacity data.

The scaling relation given by Eq. (53) clearly predicts that the distance between $T_{\mathrm{c}}$ and $T_{\mathrm{peak}}$ increases with field following a power law $H^{1/\Delta}$. Figure 14 depicts the log-log plot of $T_{\mathrm{peak}} - T_{\mathrm{c}}$ vs $H$ for **4**. It can be seen that the experimental points show a tendency to align. The solid line is the best fit whose slope is equal to the inverse of the gap exponent.

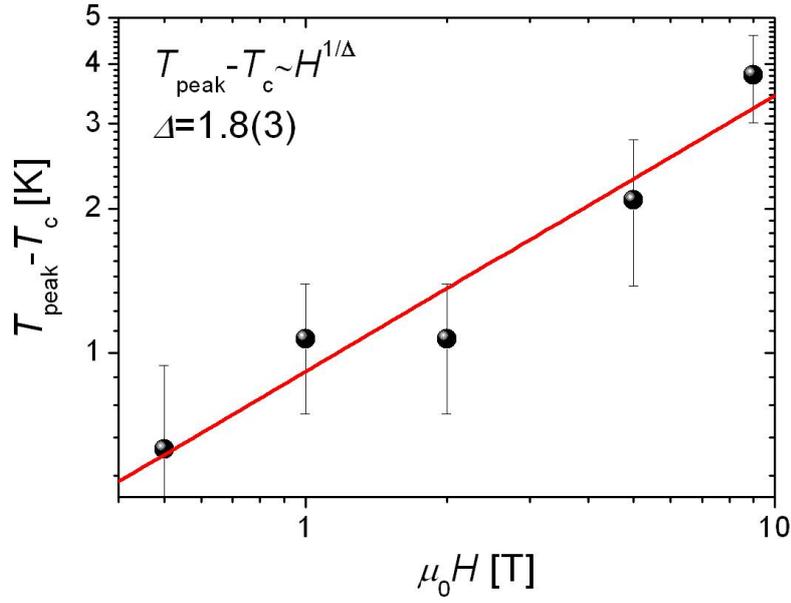

**Fig. 14.** Log-log plot of the field dependence of the distance $T_{peak} - T_c$ for **4**.

Figure 15 shows the temperature dependence of the mean exponent $n$ inferred from the $\Delta S_M$ data of Fig. 13. The curve exhibits a minimum slightly above the transition temperature $T_c = 8.3$ K. On leaving the minimum there is a relatively steep increase on the right-hand side wing towards the values exceeding 1. This behaviour is a consequence of the Curie-Weiss law holding far above $T_c$. In this temperature region the magnetization has a linear field dependence and the calculation of the magnetic entropy change using Eq. (50) leads to a quadratic field dependence of $\Delta S_M$. Hence, for the high temperature limit of the $n(T)$ curve the value of $n = 2$ is implied. The low temperature limit can be explained by realizing that well below $T_c$ and for moderate applied fields the magnetization exhibits a weak field dependence. Therefore, the integrand in Eq. (50) will be practically field independent and consequently $\Delta S_M$ will be a linear function of field, or equivalently the value of $n$ will be close to 1. The value of $n$ at the transition temperature $T_c$ is found to be equal to 0.64. This fact and the value of the gap exponent $\Delta = 1.8(3)$ imply through Eqs. (30), (33), and (56) the values of $\gamma \approx 1.4$ and $\beta \approx 0.35$ which are consistent with those predicted for the 3D Heisenberg model [26].

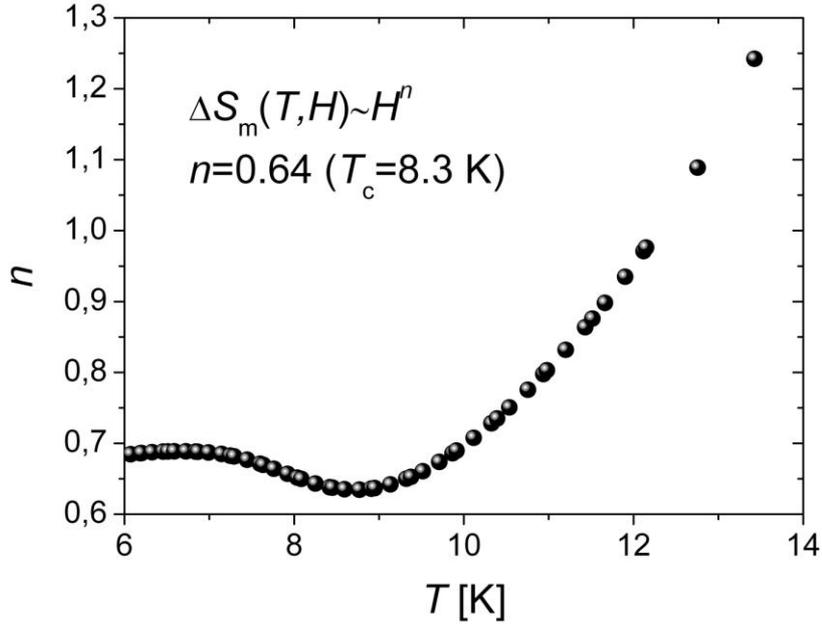

**Fig. 15.** Temperature dependence of the mean exponent $n$ inferred from the $\Delta S_M$ data in Fig. 13.

7. Combined scaling of magnetic entropy and order parameter

As was demonstrated in Section 2 the ZF µSR spectroscopy provides a direct insight into the thermal behaviour of the order parameter related to quasistatic local fields which unambiguously mark the onset of the transition to an ordered phase. This important information can be combined with that obtained from the complementary calorimetric measurements to further characterize the critical behaviour of a material. The scaling relations for specific heat and order parameter imply the validity of a combined scaling of excess entropy and the square of order parameter below the transition point $T_c$ [24]

$$\frac{\Delta S}{Q^2} \propto \left(1 - \frac{T}{T_c}\right)^{\kappa}, \qquad (57)$$

where $\kappa$ is the corresponding critical exponent related to critical exponents $\alpha$ and $\beta$,

$$\kappa = \begin{cases} 1-\alpha-2\beta & \alpha > 0, \\ 1-2\beta & \alpha \leq 0. \end{cases} \qquad (58)$$

Equivalently, this combined scaling relation can be also expressed by the following formula

$$\Delta S \propto Q^{2\kappa'}, \qquad (59)$$

where

$$\kappa' = \begin{cases} (1-\alpha)/2\beta & \alpha > 0, \\ 1/2\beta & \alpha \leq 0. \end{cases} \quad (60)$$

The excess entropy should be calculated from the excess heat capacity data starting from $T_c$ and moving towards lower temperatures, i.e.

$$\Delta S = \int_T^{T_c} \frac{\Delta C_p}{T'} dT'. \quad (61)$$

Figures 16(a) and 16(b) show the two types of the combined scaling for **1**. As the multiplicative factors are irrelevant in the scaling analysis, the thermal dependence of the order parameter was replaced by that of the local field obtained from the µSR experiment. It is apparent that the experimental points tend to align while approaching the transition temperature. The asymptotic linear behaviour implies the following values of the critical exponents, $\kappa = 0.219(3)$ and $\kappa' = 1.294(4)$. These values are close to 0.26 and 1.31, respectively, predicted for the 3D Heisenberg model [24]. Using Eqs. (58) and (60) for $\alpha < 0$ (see Fig. 5) and the value of $\beta = 0.38(1)$ found in µSR experiment, one obtains $\kappa = 0.24(2)$ and $\kappa' = 1.32(3)$, which is consistent with the values determined from the direct scaling analysis.

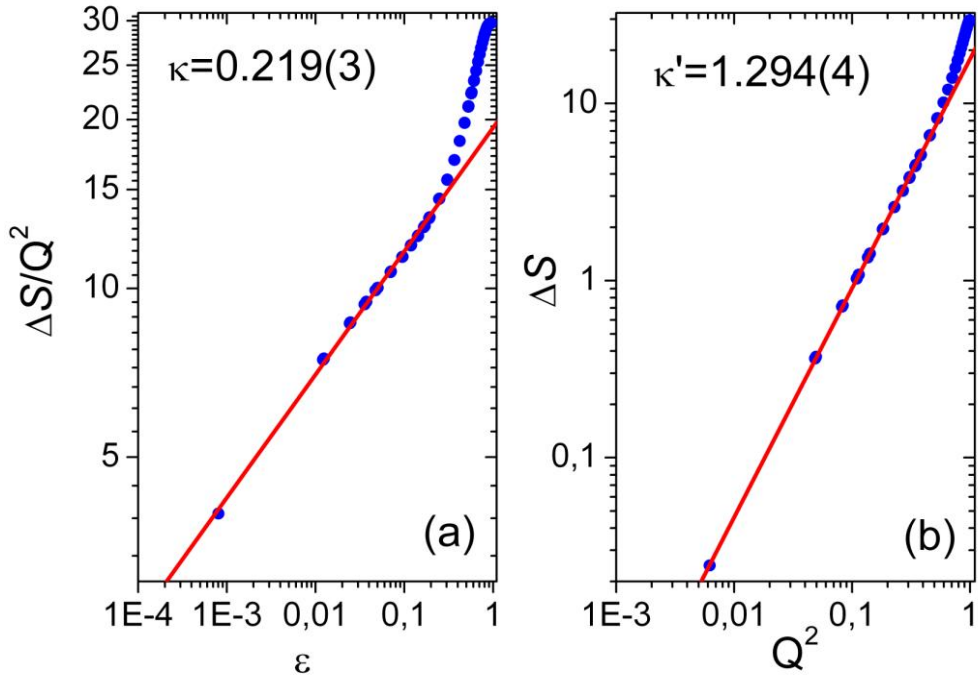

**Fig. 16.** {(a) Log-log plot of $\Delta S/Q^2$ vs $(T_c - T)/T_c$ for **1**. The slope of the best-fit line yields the exponent $\kappa$. (b) Log-log plot of $\Delta S$ vs $Q^2$ for **1**. The slope of the best-fit line gives the exponent $\kappa'$.

8. Conclusions

We have presented a wide scope of aspects of critical behaviour illustrating each with an experimental example gathered during our investigations of critical behaviour in molecular magnets showing a transition to a long-range ordered state. The included experimental data demonstrate the concept of universality; out of four molecular magnets, compounds **1**, **2**, and **4** exhibit the close affinity to the same 3D Heisenberg model, although they represent spatial arrangements determined by different space groups and consist of magnetic ions of different local structure. The scaling ideas have been shown to arise not only in the context of purely magnetic properties but also in the context of thermal properties, which was exemplified by magnetocaloric effect as well as the combined scaling of excess entropy and order parameter. Two ingenious approaches to scaling analysis were outlined, the first due to Kouvel and Fisher and the other based on the generalized Curie-Weiss law developed independently by Arrott and Souletie. The collected facts may serve as a useful introduction to scaling phenomena in magnetic materials in particular and in other condensed matter systems in general.

Acknowledgements
This work has been partially supported by Polish National Science Centre within Research Project 2011/01/B/ST5/00716.